# A novel IR-SRGAN assisted super-resolution evaluation of photothermal coherence tomography for impact damage in toughened thermoplastic CFRP laminates under room temperature and low temperature

*Pengfei Zhu[1], Hai Zhang[1,2]\*, Stefano Sfarra[3], Fabrizio Sarasini[4], Zijing Ding[5], Clemente Ibarra-Castanedo[1], Xavier Maldague[1]*

[1] Department of Electrical and Computer Engineering, Computer Vision and Systems Laboratory (CVSL), Laval University, Quebec G1V 0A6, Canada

[2] Centre for Composite Materials and Structures (CCMS), Harbin Institute of Technology, Harbin 150001, China

[3] Department of Industrial and Information Engineering and Economics (DIIIE), University of L'Aquila, I-67100 L'Aquila (AQ), Italy

[4] Department of Chemical Engineering Materials Environment & UDR INSTM, Sapienza University of Rome, Rome, I-00184, Italy

[5] School of Energy Science and Engineering,, Harbin Institute of Technology, Harbin 150001, China

\*Corresponding author: hai.zhang.1@ulaval.ca (H.Z.)

## Abstract

Evaluating impact-induced damage in composite materials under varying temperature conditions is essential for ensuring structural integrity and reliable performance in aerospace, polar, and other extreme-environment applications. As matrix brittleness increases at low temperatures, damage mechanisms shift: impact events that produce only minor delaminations at ambient conditions can trigger extensive matrix cracking, fiber/matrix debonding, or interfacial failure under severe cold loads, thereby degrading residual strength and fatigue life. Precision detection and quantification of subsurface damage features (e.g., delamination area, crack morphology, interface separation) are critical for subsequent mechanical characterization and life prediction. In this study, infrared thermography (IRT) coupled with a newly developed frequency multiplexed photothermal correlation tomography (FM-PCT) is employed to capture three-dimensional subsurface damage signatures with depth resolution approaching that of X-ray micro-computed tomography. However, the inherent limitations of IRT, including restricted frame rate and lateral thermal diffusion, reduce spatial resolution and thus the accuracy of damage size measurement. To address this, we develop a new transfer learning-based infrared super-resolution generative adversarial network (IR-SRGAN) that enhances both lateral and depth-resolved imaging fidelity based on limited thermographic datasets.

## 1. Introduction

Carbon fiber-reinforced polymer (CFRP) laminates are widely used in aerospace applications due to their high specific stiffness and strength [1,2]. However, their macroscopic mechanical response and microscopic failure mechanisms under severe cold conditions remain insufficiently understood [3,4], limiting reliable deployment in extreme environments such as high-altitude flight and polar structures. Under low-velocity impact (LVI), the brittle behavior of the polymer matrix at low temperatures can significantly alter damage initiation and progression: impacts that cause only minor delamination at ambient temperature may lead to extensive matrix cracking, fiber/matrix debonding, and interlaminar failure in low temperature regimes, thereby degrading residual load-bearing capability and fatigue life.

Mechcanically, the structural response of a laminate under impact is dominated by plate bending and dynamic load transfer across plies. [5-7]. Experimental investigations commonly employ drop-weight impact tests and quasi-static indentation (QSI) to characterize energy absorption, peak load, and damage morphology [8-10]. However, discrepancies often arise when correlating QSI results with actual impact tests due to differences in loading rate, boundary conditions, and ply stacking sequence [11-13]. This underscores the need for realistic dynamic scenarios, especially when material properties vary strongly with temperature [14].

Various non-destructive testing (NDT) techniques have been developed to evaluate the impact resistance of different composites, including ultrasound [15], infrared thermography [16], X-ray radiography [17], pulsed eddy current [18]. However, conventional ultrasound and pulsed eddy current techniques provide only limited two-dimensional information about impact damage. Due to the thin structure of composites, conventional X-ray computed tomography (CT) techniques [19] are ineffective in providing damage information along the depth direction. Although X-ray micro-CT can effectively resolve this issue, it is costly and limited to small specimens [20]. To date, no visible alternative to X-ray micro-CT has been established. Recently, advanced photothermal coherence tomography techniques have been proposed for 3D subsurface imaging [21-23]. To overcome the limitations of narrow imaging scope and high frame-rate requirements, frequency multiplexed photothermal correlation tomography (FM-PCT) was introduced [24].

In this study, the LVI damage of several CFRP and carbon fiber-reinforced thermoplastic polymer (CFRTP) composites was evaluated under low temperature (-70 °C) and room temperature conditions, including neat CFRP specimens, as well as CFRTP composites including polyetheretherketone/carbon fiber (PEEK/CF) and polyamide/carbon fiber (PA/CF) specimens. Thermal diffusivity was measured using partial time method to assess the influence of low-temperature impacts. To further evaluate the impact damage, the advanced FM-PCT technique was employed to reconstruct the 3D subsurface structure. Furthermore, to overcome the limitation of low spatial resolution, a new transfer learning-based infrared super-resolution generative adversarial networks (IR-SRGAN) was proposed for both 2D and 3D thermographic imaging. X-ray micro-computed tomography (micro-CT) was used to validate the feasibility and accuracy of the proposed 3D super-resolution photothermal coherence tomography technique.

## 2. Materials

Impact energy was applied using an instrumented drop-weight impact testing machine (CEAST/Instron 9340), equipped with a hemispherical impactor tip (12.7 mm diameter) and a total falling mass of 8.055 kg. Test specimens were pneumatically clamped between two steel plates, leaving a circular unsupported area with a diameter of 40 mm.

**Table 1**
Samples used in this work.

| | Specimens | | | |
|---|---|---|---|---|
| Experiments_1 | Neat/CF/7.5J | Neat/CF/15J | PA6.6/CF/7.5J | PA6.6/CF/15J |
| Validation (FM-PCT vs. Micro-CT) | 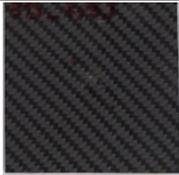 | 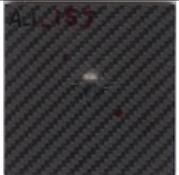 | 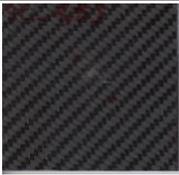 | 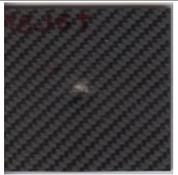 |
| Experiments_2 | PEEK/CF/15J/-70°C | PEEK/CF/5J/-70°C | PEEK/CF/15J | PEEK/CF/5J |
| Low temperature impact evaluation (FM-PCT) | 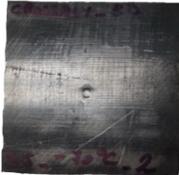 | 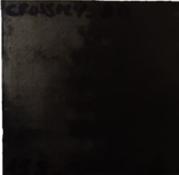 | 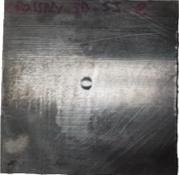 | 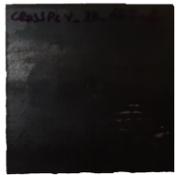 |
| Experiments_3 | PA12/CF/40J | PA12/CF/5J | PA12/CF/2.5J | |
| Room temperature impact evaluation (FM-PCT) | 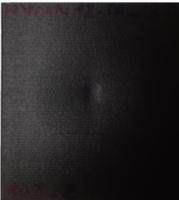 | 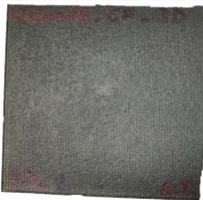 | 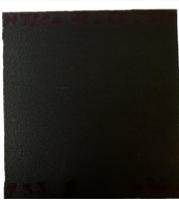 | |

*2.1. PA6.6/CF composites*

Interlayer-toughened carbon fiber-reinforced polymer (PA6.6/CF) laminates incorporating electrospun (PA6.6) veils, along with neat CFRP (Neat/CF) laminates, were fabricated by compression moulding a stack of 12 carbon fiber prepreg plies (twill 2×2, 200 g/m$^2$) at 120 °C and 4 bar for 2 hours using the VTM260 epoxy system supplied by Solvay. In the modified laminates, electrospun veils were inserted at eleven interlaminar interfaces (i.e., between adjacent carbon fiber layers) following a cross-ply stacking sequence, as shown in Table 1. The electrospun veils were provided by RevolutionFibres (Xantu.Layr®) and had an areal density of 1.5 g/m$^2$. PA6.6/CF and Neat/CF composite specimens with dimensions of 70 mm × 70 mm × 3 mm were subjected to impact energies of 7.5 J and 15 J, respectively.

*2.2. PEEK/CF composites*

PEEK/CF composites were fabricated using a hybrid composite structure comprising a polyetherketoneketone (PEEK) matrix reinforced with both short and continuous carbon fibers, as shown in Table 1. The base layer of each plate incorporated short fiber reinforcement, while the upper layers (12 in total) were reinforced with continuous carbon fibers arranged in a cross-ply (0°/90°) configuration. This layup was manufactured by 9T Labs, with alternating fiber orientation to enhance mechanical performance. The resulting laminates were consolidated by hot pressing at 450 °C to ensure strong interfacial bonding and structural integrity. These specimens are representative of high-performance thermoplastic composites suitable for demanding structural applications.

The PEEK/CF composites with dimensions of 70 mm × 70 mm × 3 mm, were subjected to impact energies of 15 J and 5 J, respectively. To investigate the temperature-dependent impact response, tests were carried out under both room temperature and low temperature conditions (−70 °C). The impact energy was controlled by varying the drop height of the impactor while maintaining a constant mass of 3.055 kg. Prior to testing, all specimens were preconditioned in a climatic chamber at the target temperature for one hour to achieve thermal equilibrium and ensure uniform temperature distribution throughout the samples.

*2.3. PA12/CF composites*

PA12/CF composites were fabricated using commercial fused deposition modeling (FDM) technology with a carbon fiber-reinforced nylon filament (Nylon 12CF), printed on a Stratasys Fortus 450mc system, as detailed in Table 1. The specimens were produced layer-by-layer with a slice height of 0.250 mm in a cross-ply (0°/90°) configuration, resulting in a total of 16 layers. Unlike the previously described group, no post-processing (e.g., compaction or thermal treatment) was applied after printing. These samples represent short fiber-reinforced thermoplastics, which are more compatible with additive manufacturing and hold promise for lightweight and customized applications. PA12/CF composites with dimensions of 70 mm × 70 mm × 3 mm were subjected to impact energies of 40 J, 5 J, and 2.5 J, respectively.

**3. Methodology**

*3.1. Frequency-multiplexed photothermal correlation tomography*

As an advanced thermophotonic imaging technique, enhanced truncated-correlation photothermal coherence tomography (eTC-PCT) not only provides depth-resolved tomographic images of both hard and soft biological tissues but also addresses the low resolution and inhomogeneous thermal perturbation issues inherent to conventional dynamic thermal tomography methods. However, its specific excitation mode—using a chirp signal—is time-consuming, and the limited imaging area (~1 cm²) significantly restricts the applicability of eTC-PCT in industrial inspection. Frequency multiplex photothermal correlation tomography (FM-PCT) combines the advantages of infrared thermography (IRT), including fast imaging speed and large imaging area, with the eTC-PCT algorithm, making it suitable for industrial inspection applications.

The FM-PCT algorithm employs general pulse excitation or linear scanning excitation to stimulate the samples, as these methods are easy to implement and well-suited for the rapid inspection of shallow defects. For one-dimensional heat diffusion, the thermal response to pulse excitation can be expressed as:

$$\tilde{T}(\xi,d,\omega) = \frac{Q(\omega)}{k\sqrt{\xi^2 + i\omega/a}} \left[ \frac{1 + \exp\left(-2d\sqrt{\xi^2 + \frac{i\omega}{a}}\right)}{1 - \exp\left(-2d\sqrt{\xi^2 + \frac{i\omega}{a}}\right)} \right] \qquad (1)$$

where $d$ denotes the defect depth, $\tilde{T}$ and $Q(\omega) = 2\tau\text{sinc}(\omega\tau/\pi)$ represent the Fourier transforms of the temperature field and heat flux, respectively. Here, $\omega$ is the temporal angular frequency in the Fourier domain, and $\xi$ is the spatial frequency with $\xi^2 = u^2 + v^2$, where $u$ and $v$ are spatial frequency variables corresponding to $x$ and $y$. $k$ is the thermal conductivity, and $\alpha$ denotes thermal diffusivity of the material. Eq. (1) provides the theoretical solution for surface temperature variation with defects in the frequency domain. Subsequently, Eq. (1) can be inverted into the spatial domain:

$$\tilde{T}(x,y,d,\omega) = \int_{-\infty}^{\infty}\int_{-\infty}^{\infty} \frac{Q(\omega)}{k\sqrt{\xi^2 + i\omega/a}} \left[ \frac{1 + \exp\left(-2d\sqrt{\xi^2 + \frac{i\omega}{a}}\right)}{1 - \exp\left(-2d\sqrt{\xi^2 + \frac{i\omega}{a}}\right)} \right] e^{i(ux+vy)} du\, dv \qquad (2)$$

The FM-PCT modality truncates the pulse excitation into multiple sinusoidal signals $\sin(2\pi f_n t)$, where $f_n = n f_s/N$ ($n$ = 1, 2, 3,…, $N/4$) and performs time-delay cross-correlation from signal $\sin(2\pi f_1 t)$ to signal $\sin(2\pi f_{n/4} t)$. As is well known, an excitation signal with lower frequency can penetrate deeper ($\mu = \sqrt{2\alpha/\omega}$, where $\mu$ is the thermal diffusion length, and $\omega$ is angular modulation frequency). Therefore, the reference signals are replaced by concurrent sinusoidal excitations ($R_0 = \cos(2\pi f_n t)$ and $R_{90} = \sin(2\pi f_n t)$) with different frequencies (from high to low).

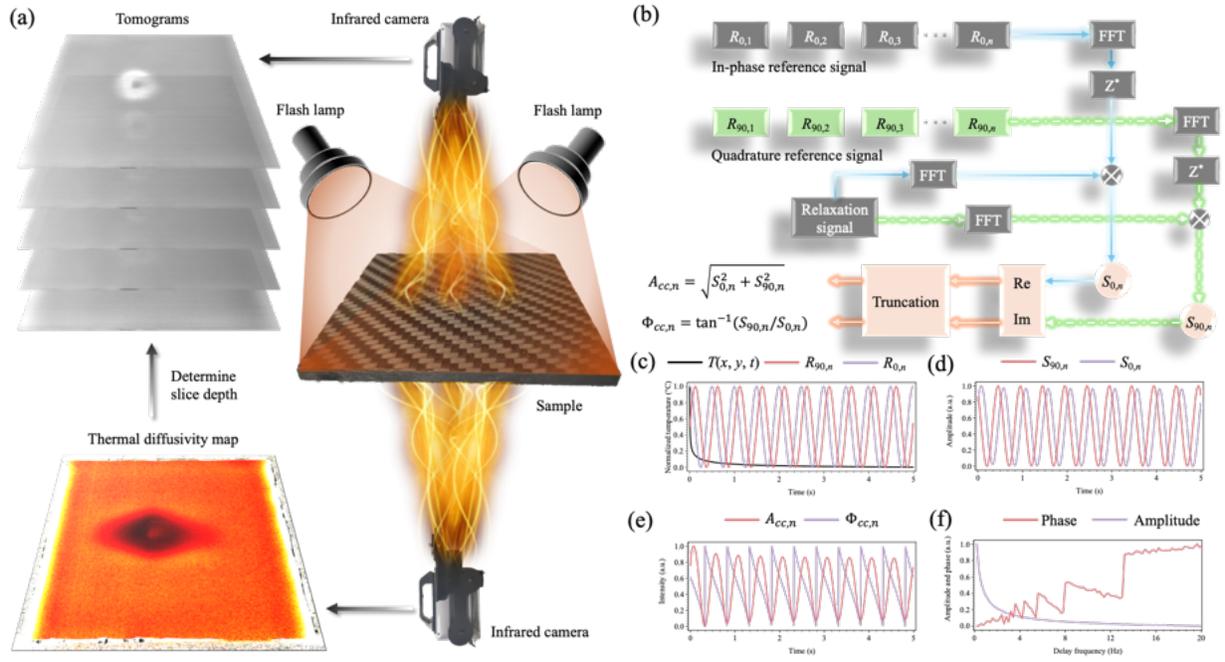

**Fig. 1.** Frequency-multiplexed photothermal coherence tomography: (a) Schematic of experimental setup based on pulsed thermography and linear scan thermography. (b) The flowchart of FM-PCT algorithm. (c) In-phase and quadrature reference signals and relaxation signals. (d) Cross-correlation amplitude. (e) Cross-correlation phase. (f) Final amplitude and phase slice information.

The cross-correlation signals with a delay component *n* can be calculated as:

$$S_{0,n} = \mathcal{F}^{-1}\left\{\frac{1}{2}\tilde{T}(x,y,\omega-\omega_n) + \frac{1}{2}\tilde{T}(x,y,\omega+\omega_n)\right\} \quad (3)$$

$$S_{90,n} = \mathcal{F}^{-1}\left\{-\frac{1}{2i}\tilde{T}(x,y,\omega-\omega_n) + \frac{1}{2i}\tilde{T}(x,y,\omega+\omega_n)\right\} \quad (4)$$

where $\omega_n = 2\pi f_n$. Finally, the amplitude ($A_{CC}$) and phase ($\Phi_{CC}$) can be given as:

$$A_{cc,n} = \sqrt{S_{0,n}^2 + S_{90,n}^2} \tag{5}$$

$$\Phi_{cc,n} = \tan^{-1}(S_{90,n}/S_{0,n}) \tag{6}$$

The schematic of the experimental setup is shown in Fig. 1(a), and the signal processing flowchart of the FM-PCT algorithm is illustrated in Fig. 1(b). The in-phase ($R_0$) and quadrature ($R_{90}$) reference signals are synthesized from the excitation pulse, as shown in Fig. 1(b). The resulting time-delay measurement determines the depth of the absorber. Cross-correlation signals ($S_0$ and $S_{90}$) are generated by matched filtering between the infrared response signal and the reference signals. The amplitude cross-correlation ($A_{cc}$) and phase cross-correlation ($\Phi_{cc}$) are calculated using Eq. (5) and Eq. (6), respectively, and subsequently truncated at consecutive time intervals using a time-gating filter to extract depth-resolved amplitude and phase information, as shown in Fig. 1(b).

### 3.2. Transfer learning-based super-resolution networks

In infrared thermography and photothermal coherence tomography, the frame rate plays a critical role in determining depth resolution, as it is directly linked to thermal diffusion length ($\mu = \sqrt{2\alpha/\omega}$). Furthermore, frame rate can also affect lateral resolution. This is because the trade-off between frame rate and spatial resolution in IRT mainly stems from limitations in data bandwidth and detector readout speed. As frame rate increases, the infrared camera must capture, process, and transmit more frames per second, which significantly increases data throughput. Without reducing image size, this can lead to delays and dropped frames. Although frequency-multiplexed photothermal correlation tomography (FM-PCT) can reconstruct three-dimensional subsurface information from a limited number of frames, its depth resolution inevitably decreases. To address this limitation, we propose an infrared super-resolution generative adversarial network (IR-SRGAN) to enhance both lateral and depth resolution in FM-PCT.

It is well known that publicly available datasets for infrared thermography are limited, particularly for high-resolution training. However, the performance of super-resolution tasks heavily relies on large-scale, high-quality datasets. Directly applying super-resolution models trained on visible-spectrum images to infrared thermograms often results in unrealistic textures and artifacts due to the modality gap. To overcome these limitations, we propose, for the first time, a transfer learning-based super-resolution generative adversarial network tailored specifically for infrared thermography.

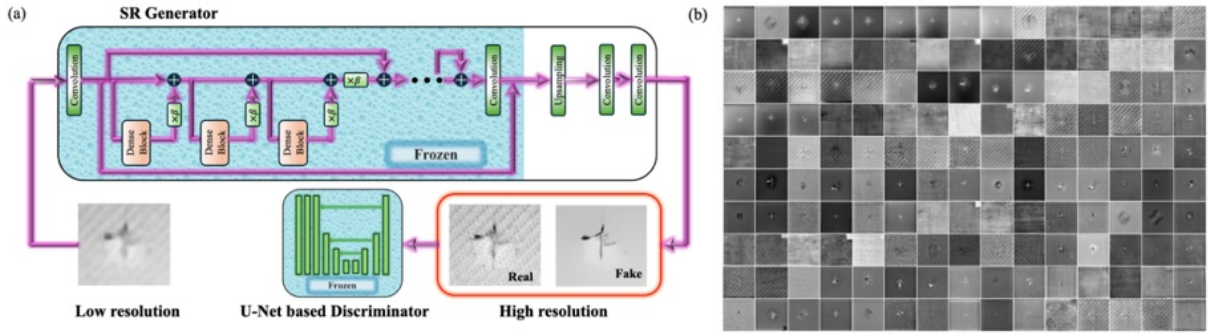

**Fig. 2.** (a) The schematic of transfer learning-based infrared super-resolution generative adversarial networks (IR-SRGAN). (b) Part of the training datasets.

As illustrated in Fig. 2(a), the input thermogram is first processed by the generator network of IR-SRGAN. Initially, the thermogram passes through a convolutional layer, followed by a stack of 23 Residual-in-Residual Dense Blocks (RRDBs) [25], which effectively integrate multi-level residual learning with dense connectivity to enhance feature representation. After traversing the RRDB modules, the super-resolved image is reconstructed through a sequence of three convolutional layers and a final upsampling module. These components together constitute the overall architecture of the generator.

After processing by the SR generator, the resulting super-resolution (SR) images are evaluated using a U-Net-based discriminator [26]. This discriminator is trained to differentiate between real high-resolution images from the training dataset and synthetic images generated by the generator. As shown in Fig. 2, typical artifacts distinguishing fake from real images include blurring and unnatural edge transitions. The encoder module $D_{enc}^U$ extracts hierarchical representations, while the decoder module $D_{dec}^U$ enhances spatial resolution through pixel-wise classification. This architecture allows the discriminator to access both global image realism and local structural fidelity. The discriminator loss is calculated by aggregating the classification outcomes from both $D_{enc}^U$ and $D_{dec}^U$:

$$\mathcal{L}_{D^U} = \mathcal{L}_{D_{enc}^U} + \mathcal{L}_{D_{dec}^U} \tag{7}$$

$$\mathcal{L}_{D_{enc}^U} = -\mathbb{E}_x[\log D_{enc}^U(x)] - \mathbb{E}_z[\log(1 - D_{enc}^U(G(z)))] \tag{8}$$

$$\mathcal{L}_{D_{dec}^U} = -\mathbb{E}_x\left[\sum_{i,j} \log[D_{dec}^U(x)]_{i,j}\right] - \mathbb{E}_z\left[\sum_{i,j} \log(1 - [D_{dec}^U(G(z))]_{i,j})\right] \tag{9}$$

where $G$ aims to map a latent variable $z \sim p(z)$ sampled from a prior distribution to a realistic-looking image, while $D$ aims to distinguish between real $x$ and generated $G(z)$ images. $[D_{dec}^U(x)]_{i,j}$ and $[D_{dec}^U]_{i,j}$ refer to the discriminator decision at pixel $(i,j)$. Correspondingly, the generator objective becomes:

$$\mathcal{L}_G^{adv} = -\mathbb{E}_z[\log D_{enc}^U(G(z))] - \mathbb{E}_z\left[\sum_{i,j} \log[D_{dec}^U(G(z))]_{i,j}\right] \tag{10}$$

To overcome drawbacks of the sparse activated features and inconsistent reconstructed brightness, the perceptual loss $L_{percep}$ proposed by Wang [25] is chosen.

$$L_G = L_{percep} + \lambda L_G^{adv} + \eta L_1 \tag{11}$$

where $L_1 = \mathbb{E}_{I^{LR}}||G(I^{LR}) - y||_1$ is the content loss that evaluates the L₁-norm distance between recovered image $G(I^i)$ and the ground-truth $y$, and $\lambda, \eta$ are the coefficients to balance different loss terms.

To address the challenge of high time complexity, a fine-tuning strategy based on transfer learning [27] was employed, as illustrated in Fig. 2. Transfer learning allows the use of a pre-trained model from a related task to reduce training time and enhance performance, particularly when training data is scarce or computational resources are limited. In this study, the training dataset was compiled from Refs. [28-30], consisting of 165 high-resolution images and their corresponding 165 low-resolution counterparts, as shown in Fig. 2(b). The low-resolution images were generated by intentionally removing pixels to simulate image degradation. During fine-tuning, the discriminator was frozen to avoid introducing unnecessary noise, as it had already been adequately trained on visible-spectrum images. In contrast, the upsampling module and the last two convolutional blocks of the generator were retrained. This decision is based on the understanding that the early layers of the generator mainly extract low-level, domain-invariant features (e.g., edges and textures), which are generally transferable across tasks. However, the deeper layers—particularly those involved in upsampling and image reconstruction—are more task-specific and therefore benefit from adaptation to the new dataset.

**4. Experimental setup**

Fig. 3 shows the experimental setup of pulsed thermography (PT). A cooled infrared camera (FLIR X8501sc, 3-5 μm, InSb, NedT <20 mK, 1280 × 1024 pixels) and two Xenon flash lamps (Balcar, 6.4 kJ for each, 2 ms) have been used.

To validate the accuracy of 3D reconstruction achieved by the photothermal coherence tomography technique, micro-computed tomography (micro-CT) was used. In the micro-CT setup, X-rays are generated by a microfocus X-ray source (Skyscan1172, μCT system, Bruker MicroCT, Kontich, Belgium), transmitted through the sample, and captured as 2D projection images by a flat-panel detector. The sample is incrementally rotated on a precision rotational stage, and successive X-ray projections are acquired over a 180° rotation. A 3D model is then constructed using an inverse Radon transform algorithm.

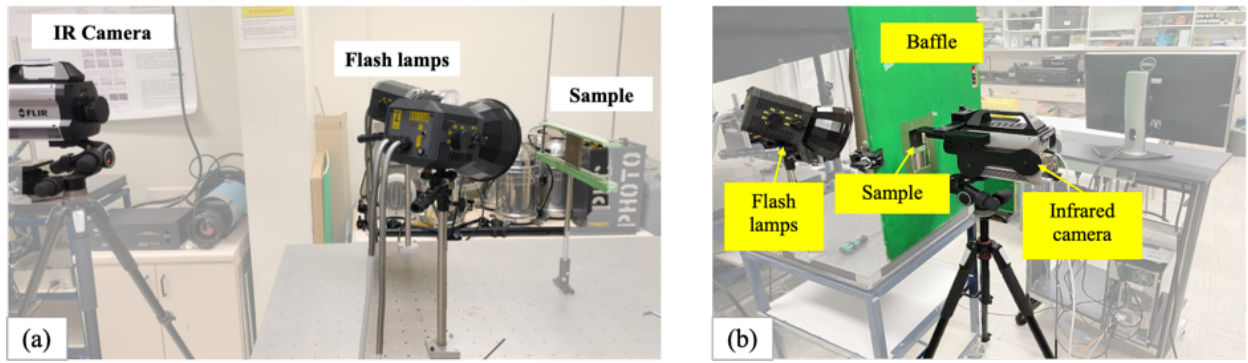

**Fig. 3.** Experimental setup of pulsed thermography in reflection mode (a) and transmission mode (b).

## 5. Results and discussion

*5.1. Experimental validation between FM-PCT and X-ray Micro-CT*

As a novel photothermal tomography technique, FM-PCT exhibits strong potential for non-destructive evaluation. To assess the accuracy of FM-PCT, X-ray micro-CT was employed to inspect Neat/CF and PA6.6/CF composite laminates subjected to varying impact energies. The comparative results obtained from FM-PCT and X-ray micro-CT are presented in Fig. 4. The depth range captured by FM-PCT was found to be 0.16 mm to 1.59 mm for Neat/CF, and 0.15 mm to 1.53 mm for PA6.6/CF. These values were calculated using the equation $z = 1.8\sqrt{\alpha/(\pi f)}$, where $\alpha$ is the thermal diffusivity (Neat/CF: 4.90×10$^{-7}$ m²/s, PA6.6/CF: 4.53×10$^{-7}$ m²/s), as measured in Ref. [28]. The frequency range used in FM-PCT analysis extends from $f_s/N$ to 100*$f_s/N$, where the frame rate $f_s$ = 150 Hz and the total number of frames $N$ = 749.

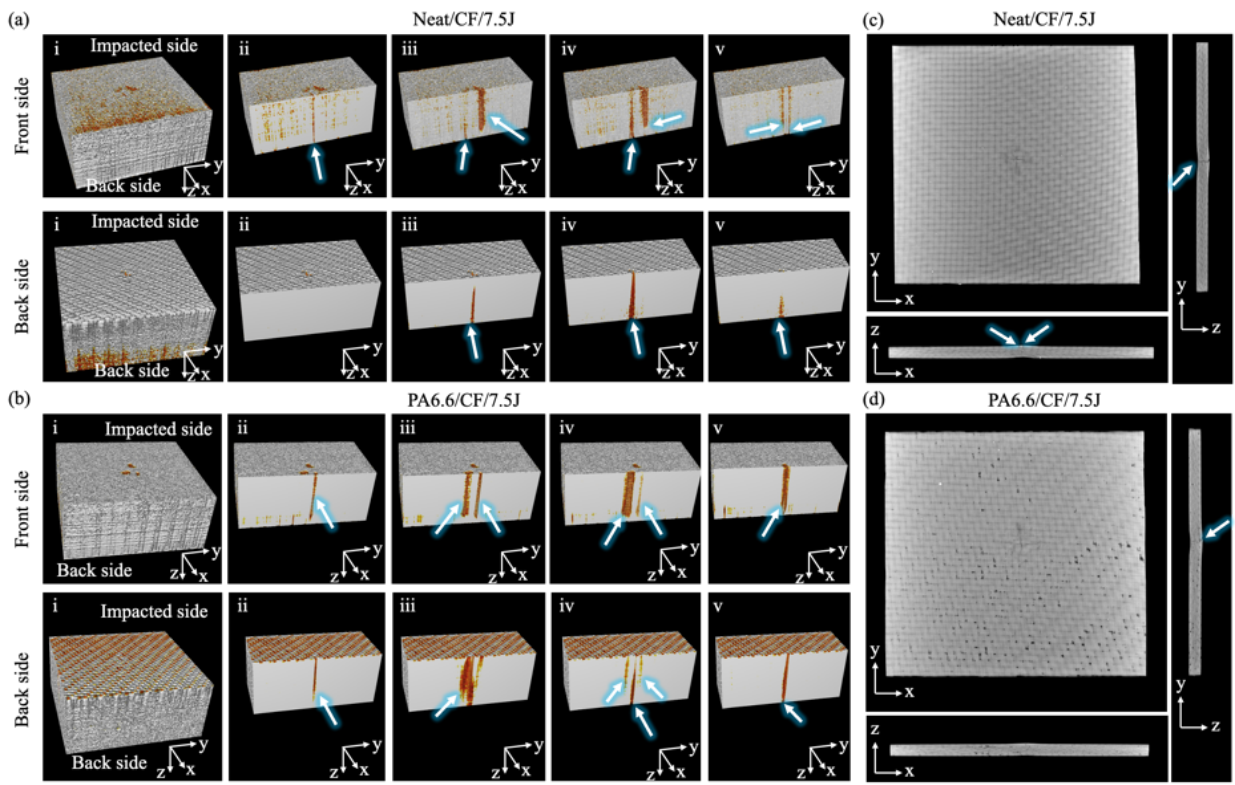

**Fig. 4.** Comparison between FM-PCT and X-ray micro-CT: (a) and (b) demonstrate tomograms of Neat/CF and PA6.6/CF subjected to 7.5 J impact energy where front side and back side denote tested surface by IRT. (c) and (d) demonstrate tomograms of X-ray micro-CT.

As shown in Fig. 4(a), two fiber fractures are observed on the front surface of the Neat/CF/7.5J sample, which correspond well to the fiber breakage seen in the z-x plane of the X-ray micro-CT results (see Fig. 4(c)). Additionally, a crack on the rear surface of the same laminate is visible in Fig. 4(a), which aligns with the damage observed in the z-y plane of the micro-CT image (see Fig. 4(c)). For the PA6.6/CF/7.5J laminate, FM-PCT reveals two prominent fiber fractures on the front side and three on the rear side (see Fig. 4(b)). However, corresponding damage features in the micro-CT images (Fig. 4(d)) are less apparent, with only one shallow crack visible in the y-z plane. Delamination is not clearly observed in the micro-CT results, highlighting the superior sensitivity of FM-PCT in detecting subsurface fiber breakage and potential interlaminar defects.

The results for Neat/CF and PA6.6/CF composites subjected to an impact energy of 15 J are shown in Fig. 5. The depth ranges reconstructed by FM-PCT are 0.16 mm to 1.59 mm for Neat/CF and 0.15 mm to 1.53 mm for PA6.6/CF. These values were calculated using the frequency range from $f_s/N$ to $100*f_s/N$, where the frame rate $f_s$ = 150 Hz and the total number of frames $N$ = 749. For the Neat/CF/15J laminate, a deep central crack is clearly observed in the FM-PCT image (see Fig. 5(a)), which corresponds well to the damage identified in the X-ray micro-CT results (see Fig. 5(c)). Notably, no signs of delamination are found in the Neat/CF/15J laminate. In contrast, the PA6.6/CF/15J laminate exhibits two delaminations on the front surface, as shown in Fig. 5(b). These features are consistent with those revealed in the z-x view of the micro-CT image (Fig. 5(d)). Additionally, the surface crack in PA6.6/CF/15J is noticeably larger than that observed in Neat/CF/15J, indicating more severe damage. Taken together, the results in Fig. 4 and Fig. 5 demonstrate that Neat/CF composites exhibit greater resistance to impact-induced damage compared to PA6.6/CF composites.

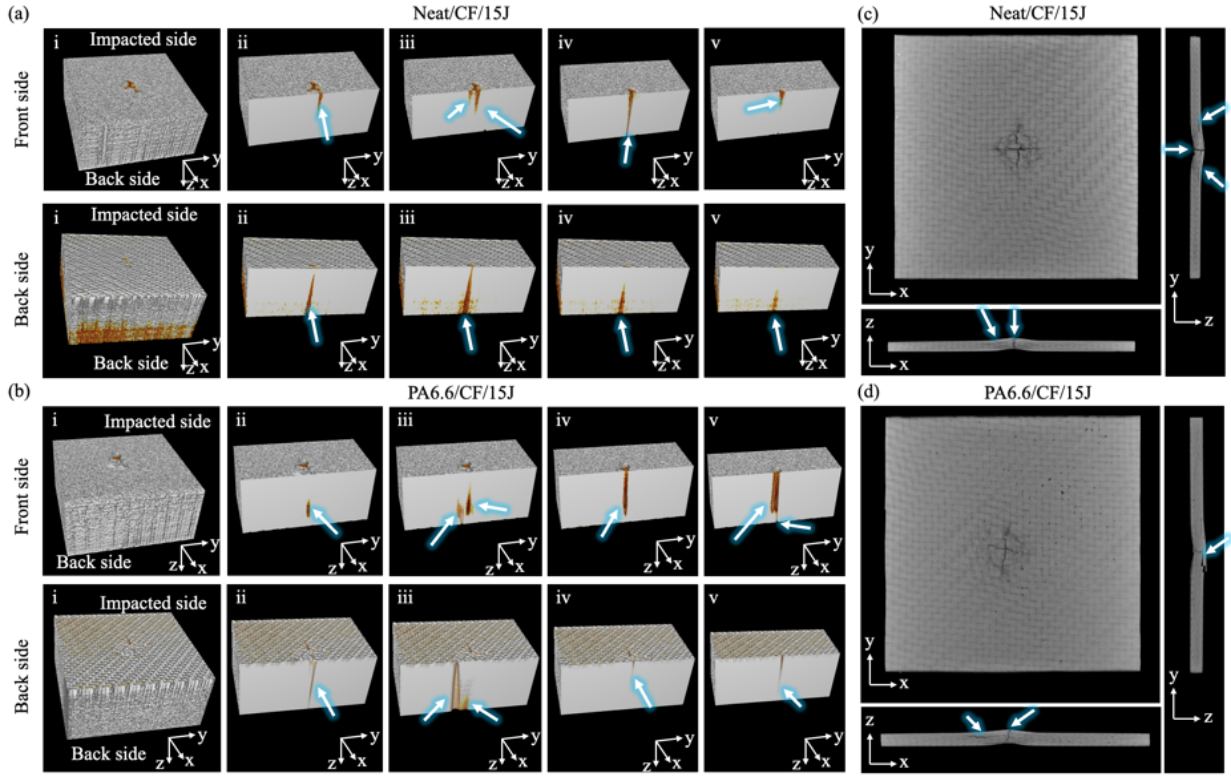

**Fig. 5.** Comparison between FM-PCT and X-ray micro-CT: (a) and (b) demonstrate tomograms of Neat/CF and PA6.6/CF subjected to 15 J impact energy where front side and back side denote tested surface by IRT. (c) and (d) demonstrate tomograms of X-ray micro-CT.

*5.2. Evaluation of low velocity impact under low temperature*

Since the feasibility and accuracy of FM-PCT were validated in the previous section, it is now possible to quantitatively evaluate impact damage under low temperatures. However, thermal diffusivity is a crucial parameter for quantifying mechanical response and defect depths. According to partial time method [31,32], the thermal diffusivity can be calculated using the flash pulse method in transmission mode. Figure 6 shows the thermal diffusivity maps of PEEK/CF composites calculated by partial time method, where the image of 4th column represents the average values of the three images on the left. It is evident that thermal diffusivity decreases at low temperatures. This reduction is attributed to changes in the material's mechanical and interfacial properties. Specifically, at low temperatures, the polymer matrix becomes more brittle, and the thermal expansion mismatch between the fibers and matrix increases. These effects make the composite more prone to microcracking and interfacial debonding during impact. Such damage mechanisms disrupt heat conduction pathways within the composite, leading to a decrease in thermal conductivity. This behavior can be described by the Maxwell-Eucken model [33]:

$$k_{eff} = k_m \left[\frac{2k_m + k_c - 2\varepsilon(k_m - k_c)}{2k_m + k_c + \varepsilon(k_m - k_c)}\right] \quad (18)$$

where $k_m$ is the thermal conductivity of the matrix, $k_c$ is the thermal conductivity of cracks/delaminations (often approximated as 0), $\varepsilon$ is the volume fraction of cracks/delaminations. When cracks or delaminations increase due to low-velocity impact, the volume fraction $\varepsilon$ increases, resulting in a significant drop in $k_{eff}$, and thus reduced thermal diffusivity ($\alpha = k_{eff}/\rho c$ where $\rho$ is the density and $c$ is the heat capacity).

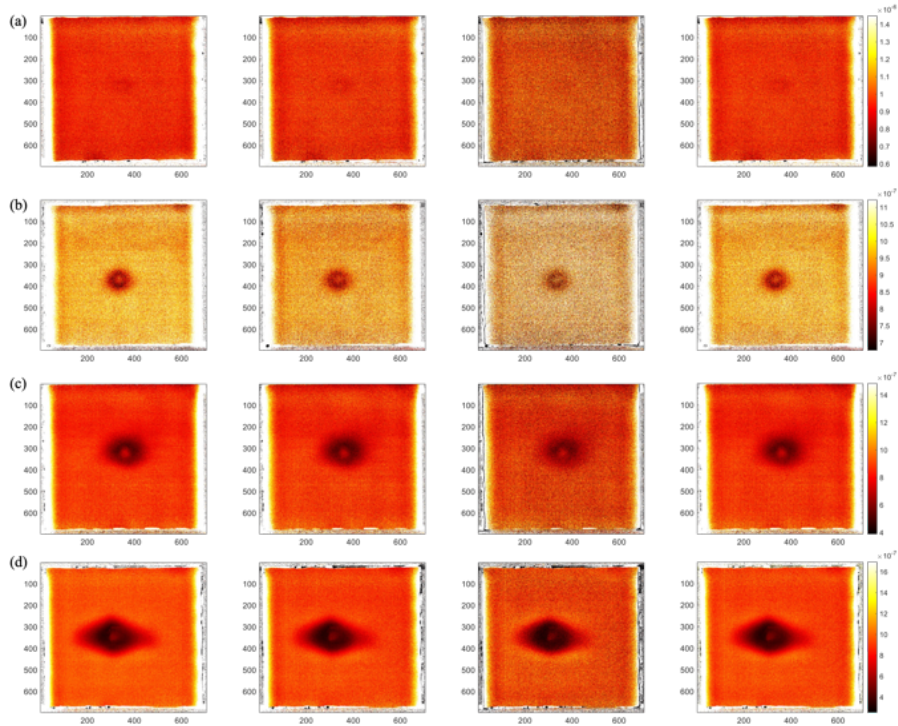

**Fig. 6.** Thermal diffusivity map: (a) PEEK/CF/5J. (b) PEEK/CF/5J/-70°C. (c) PEEK/CF/15J. (d) PEEK/CF/15J/-70°C. The images of columns 1-3 was calculated by partial time method [31,32]. The image of 4th column is the average values from columns 1-3.

To quantitatively analyze the variation of thermal diffusivity, values were extracted from both sound and damaged areas, as summarized in Table 2. It is evident that the thermal diffusivity significantly decreases when low-velocity impacts (LVIs) are conducted under low temperatures. Moreover, this reduction becomes more pronounced with increasing impact energy.

**Table 2**
Thermal diffusivity of PEEK/CF composites ($\Delta$ is the thermal diffusivity difference between sound area and damaged area, and the unit is $1\times10^{-7}$ m$^2$/s).

| Theory | Thermal diffusivity (sound area) | | | | Thermal diffusivity (damaged area) | | | | |
|---|---|---|---|---|---|---|---|---|---|
| | Eq. (15) | Eq. (16) | Eq. (17) | Mean | Eq. (15) | Eq. (16) | Eq. (17) | Mean | $\Delta$ |
| PEEK/CF/5J | 9.43 | 9.47 | 9.37 | 9.42 | 9.14 | 9.23 | 9.41 | 9.26 | 0.16 |
| PEEK/CF/5J/-70°C | 9.85 | 9.63 | 9.90 | 9.79 | 8.02 | 8.12 | 8.18 | 8.11 | 1.68 |
| PEEK/CF/15J | 8.57 | 8.47 | 8.22 | 8.42 | 5.92 | 5.82 | 6.03 | 5.92 | 2.50 |
| PEEK/CF/15J/-70°C | 9.54 | 8.99 | 8.66 | 9.07 | 4.29 | 4.10 | 3.87 | 4.08 | 4.99 |

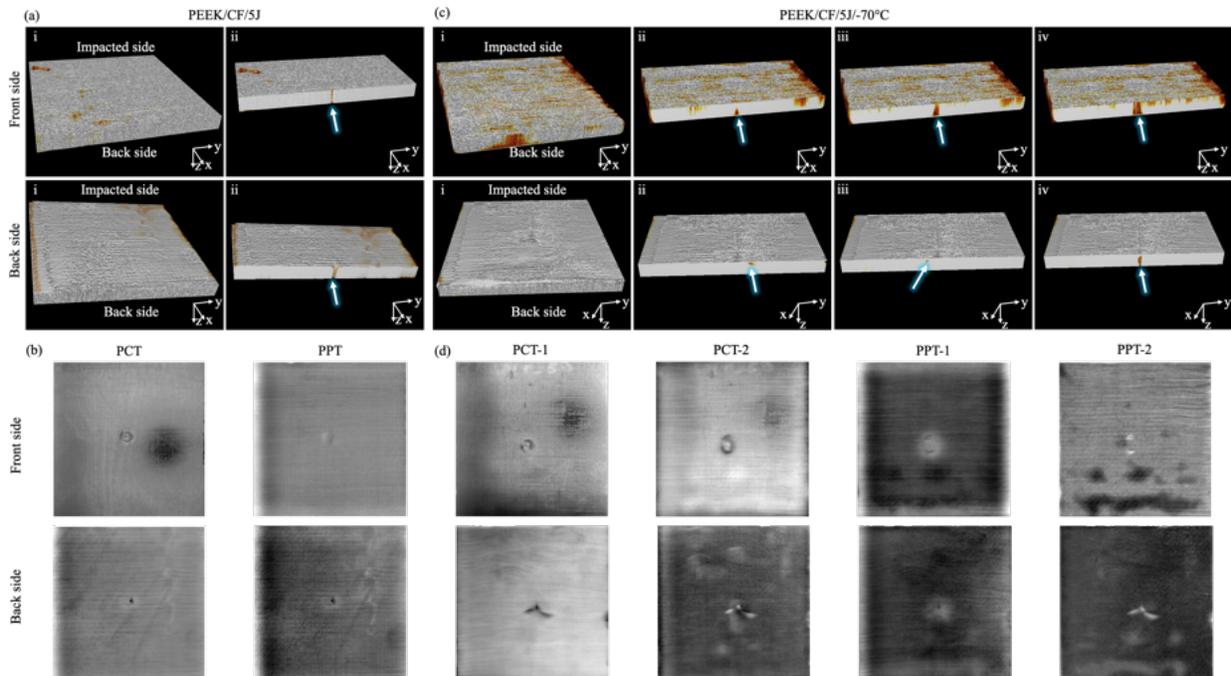

**Fig. 7.** Evaluation of PEEK/CF/5J under room and low temperatures: (a) tomograms of PEEK/CF laminate at room temperature on both front and back sides. (b) image processing based PCT and PPT algorithms on both front and back sides. (c) tomograms of PEEK/CF/5J laminate at -70°C temperature on both front and back sides. (d) image processing based PCT and PPT algorithms on both front and back sides.

After measuring the thermal diffusivity, the slice depth in FM-PCT can be calculated. The tomograms of PEEK/CF/5J and PEEK/CF/5J/-70°C composites are illustrated in Fig. 7. The depth range of FM-PCT results extends from 0.50 mm to 3 mm for PEEK/CF/5J, and from 0.55 mm to 3 mm for PEEK/CF/5J/-70°C, where the frequency range spans from $f_s/N$ to $55*f_s/N$ (with $f_s$ = 70 Hz as the frame rate and $N$ = 999 as the total number of frames). Two classic image processing algorithms, pulsed phase thermography (PPT) [34] and principal component thermography (PCT) [35], were applied to extract damage features. However, evaluating impact damage based solely on these limited 2D results remains challenging, despite the use of PPT and PCT methods. In contrast, FM-PCT provides 3D information of subsurface structures, enabling more accurate assessment of impact damage. According to tomograms on the front side (tomogram (ii) in Fig. 7(a)), a 0.78 mm crack is observed in PEEK/CF/5J, whereas the crack size increases to 3.36 mm in PEEK/CF/5J/-70°C. On the back side, the crack size measures 2.09 mm in PEEK/CF/5J. Additionally, for PEEK/CF/5J/-70°C, besides fiber fractures, two interlayer delaminations measuring 2.87 mm are detected.

The tomograms of PEEK/CF/15J and PEEK/CF/15J/-70°C composites are illustrated in Fig. 8. The depth range of FM-PCT results spans from 0.53 mm to 3 mm for PEEK/CF/5J and from 0.55 mm to 3 mm for PEEK/CF/5J/-70°C. The corresponding frequency range is from $f_s/N$ to $55*f_s/N$, where $f_s$ = 70 Hz is the frame rate, and $N$ = 999 is the total number of frames. Cross-sectional tomograms along the $z$ and $x$ directions are shown in Fig. 8. For the PEEK/CF/15J laminate, only several fiber fractures are observed (see Fig. 8(a)). In contrast, LVIs at low temperature induce not only cracks and delaminations (as seen in the tomograms along the *x* direction in Fig. 8(b)) but also matrix fractures (as shown in the *z*-direction tomograms in Fig. 8(b)) in the PEEK/CF/15J/-70°C composite. The measured crack length is 6.31 mm in PEEK/CF/15J and increases to 7.99 mm in PEEK/CF/15J/-70°C.

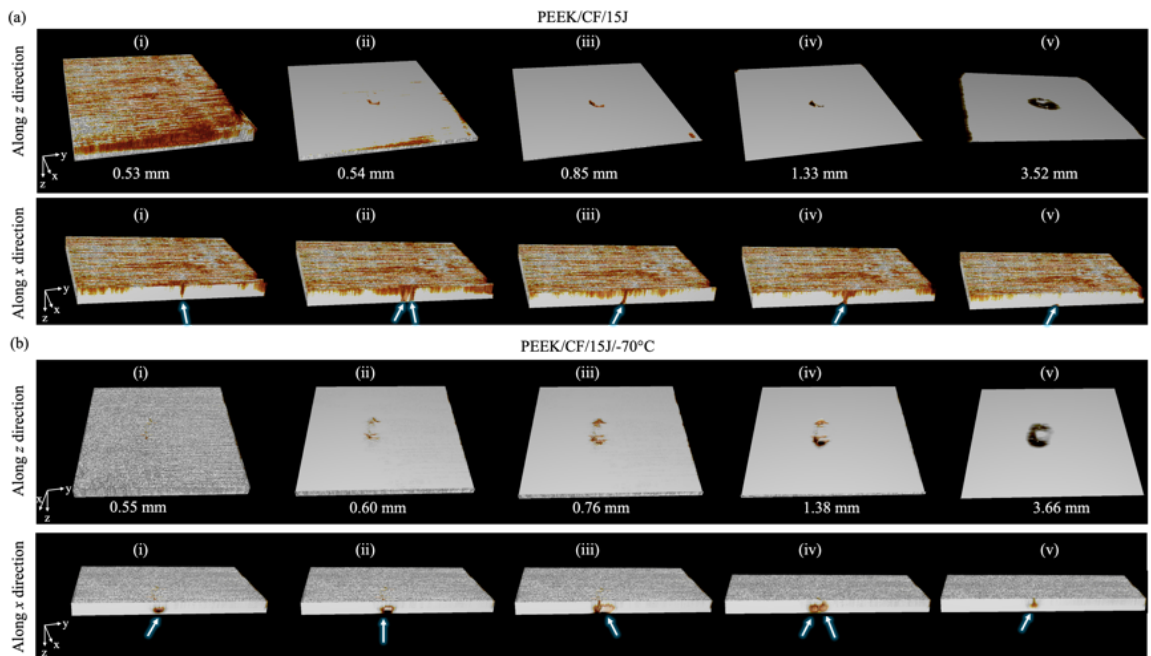

**Fig. 8.** Evaluation of PEEK/CF/15J under room and low temperatures: (a) tomograms of PEEK/CF composite at room temperature along *z* and *x* directions. (b) tomograms of PEEK/CF composite at -70°C temperature along *z* and *x* directions.

The tomograms of PA12/CF laminates tested at room temperature are presented in Fig. 9. As shown in Fig. 9(a), no damage is observed in the PA12/CF laminate under an impact energy of 2.5 J. When the impact energy increases to 5 J, a surface dent appears on the front side, and a crack is visible on the back side (Fig. 9(b)). At 40 J impact energy, two orthogonal cracks are observed on the back side (Fig. 9(c)). Additionally, four deep cracks are identified on the back side, as shown in Fig. 9(c) and (d). Notably, no delamination is detected in the PA12/CF laminates, even under the highest impact energy of 40 J. Compared with the results obtained for PEEK/CF (Fig. 8(a)) and PA6.6/CF (Fig. 5(b)), the

PA12/CF laminates exhibit different modalities of impact energy absorption (damage mode).

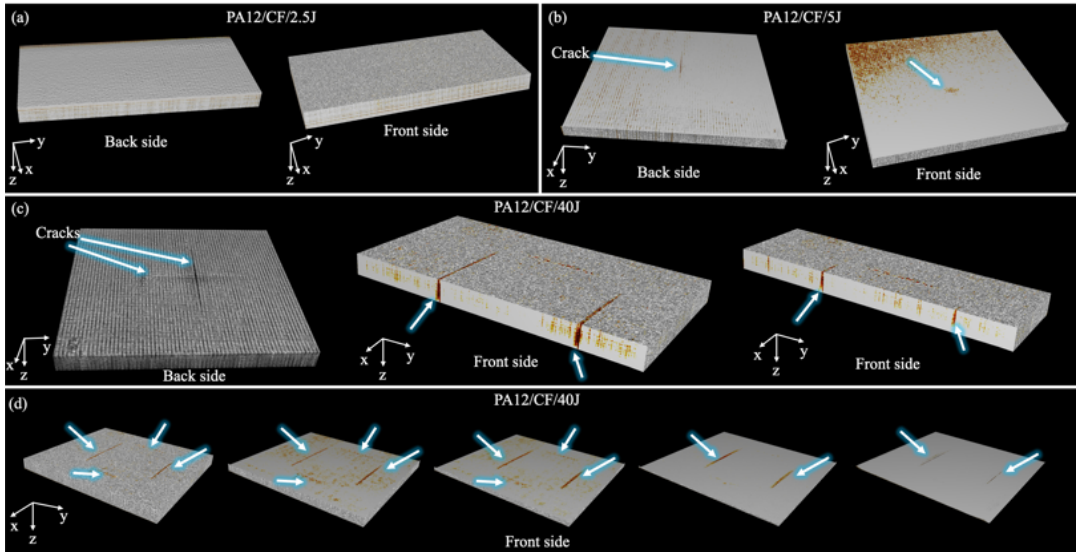

**Fig. 9.** Evaluation of PA12/CF under room temperature: (a) tomograms of PA12/CF/2.5J laminate. (b) tomograms of PA12/CF/5J laminate. (c) tomograms of PA12/CF/40J laminate. (d) tomograms of PA12/CF/40J laminate along *z* direction.

*5.3. Transfer learning-based super-resolution networks*

In infrared thermography (IRT) and photothermal coherence tomography, although advanced infrared cameras such as the FLIR SC6000 offer a spatial resolution of 640×512 pixels (~0.56 mm), they still fall short in detecting micro-delaminations and micro-cracks. Moreover, inevitable lateral thermal diffusion significantly degrades thermogram quality, particularly when detecting subsurface defects at greater depths. Deep learning has emerged as a promising technique for achieving super-resolution imaging with minimal hardware cost. However, the availability of large-scale annotated datasets in the field of IRT, especially for composite material evaluation, remains limited. Transfer learning offers an effective solution in such data-constrained scenarios by leveraging knowledge from related domains. In this study, we demonstrate, for the first time, the successful application of transfer learning in infrared thermography and photothermal tomography for the evaluation of composite materials.

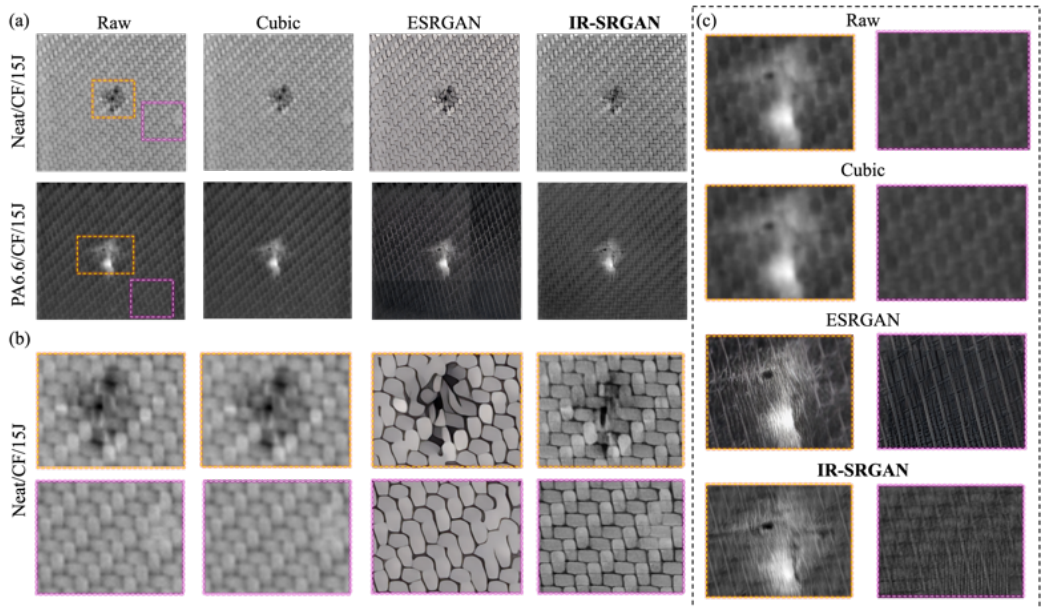

**Fig. 10.** Super-resolution imaging for Neat/CF/15J and PA6.6/CF/15J using cubic interpolation, ESRGAN, and the proposed IR-SRGAN method: (a) overall images. (b) local details of Neat/CF/15J. (c) local details of PA6.6/CF/15J. Yello box is used to observe damage, and pink box is used to observe texture.

The super-resolution imaging results for Neat/CF/15J and PA6.6/CF/15J are presented in Fig. 10. Three methods were employed to enhance spatial resolution: cubic interpolation, ESRGAN, and the proposed IR-SRGAN. To facilitate detailed observation, yellow and pink boxes were used to highlight specific regions in the original images—where the yellow box indicates the damage area and the pink box highlights the composite texture. As shown in Fig. 10(b), cubic interpolation results in a blurrier image, failing to recover fine details. Although ESRGAN improves spatial resolution, it introduces unnatural textures due to the mismatch between the training dataset and the inference images. By incorporating a transfer learning strategy, IR-SRGAN effectively reconstructs natural textures in both damaged and undamaged regions. For PA6.6/CF/15J, as shown in Fig. 10(a) and (c), ESRGAN enhances resolution and produces natural textures;

however, inconsistencies in brightness (contrast) are observed in two regions in Fig. 10(a). Additionally, the twill weave pattern in the lower-right region of the PA6.6/CF/15J image is not fully restored.

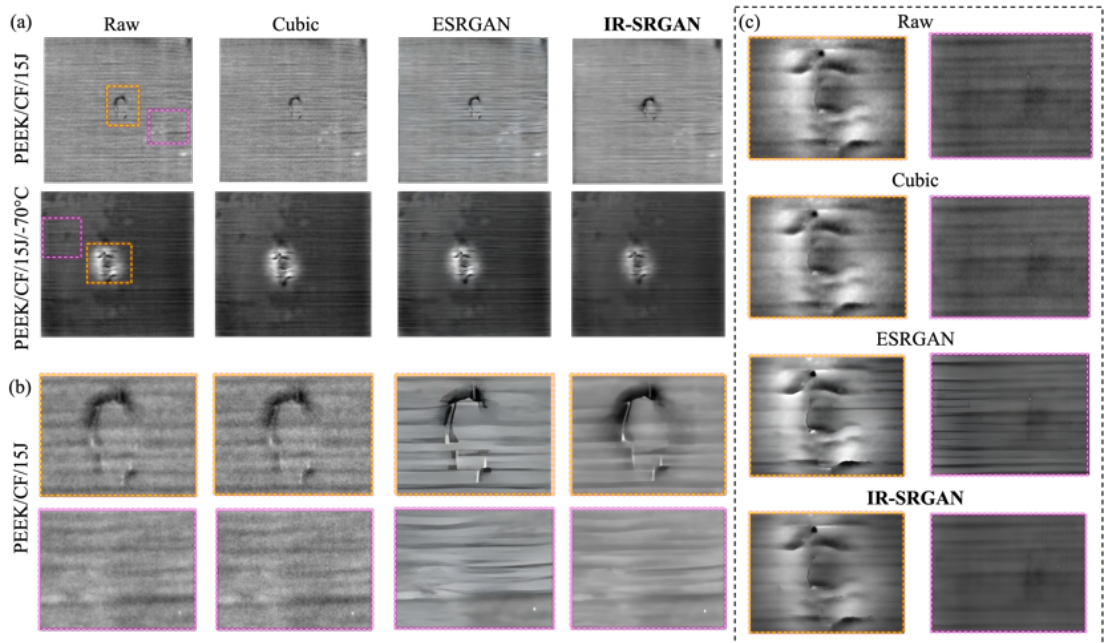

**Fig. 11.** Super-resolution imaging for PEEK/CF/15J and PEEK/CF/15J/-70°C using cubic interpolation, ESRGAN, and the proposed IR-SRGAN method: (a) overall images. (b) local details of PEEK/CF/15J. (c) local details of PEEK/CF/15J/-70°C. Yello box is used to observe damage, and pink box is used to observe texture.

The super-resolution results of PEEK/CF/15J specimens under both room and low temperatures are shown in Fig. 11. As illustrated in Fig. 11(b), although ESRGAN enhances image contrast more effectively than the other methods, it generates unnatural textures in the damaged region. A similar issue is observed in the undamaged region in Fig. 11(c), where the texture appears distorted. In contrast, the proposed IR-SRGAN not only enhances spatial resolution but also preserves natural texture features in both damaged and undamaged regions.

The results of PA12/CF subjected to 40 J impact energy are presented in Fig. 12. It is observed that ESRGAN tends to eliminate texture features in the undamaged regions. In contrast, the proposed IR-SRGAN is able to effectively reconstruct the characteristic 0°/90° weave pattern, preserving the structural details of the composite.

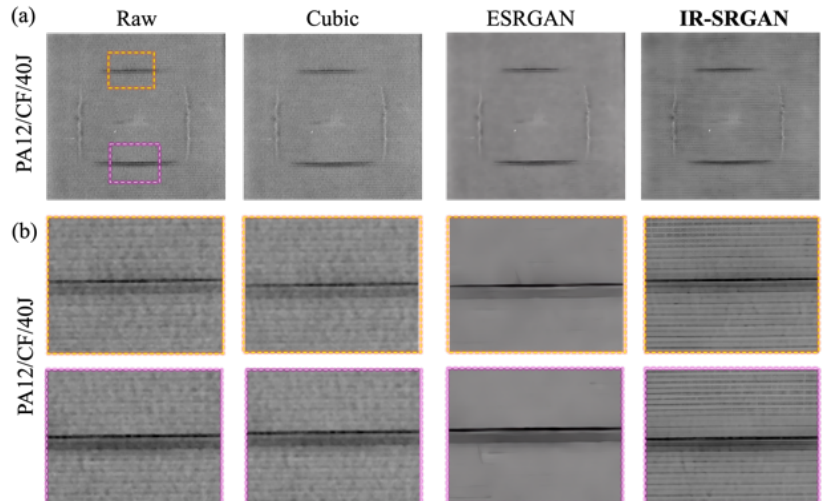

**Fig. 12.** Super-resolution imaging for PA12/CF/40J using cubic interpolation, ESRGAN, and the proposed IR-SRGAN method: (a) overall images. (b) local details of PA12/CF/40J.

## 6. Conclusion

Low-velocity impact (LVI) under low temperature conditions exhibits distinct failure mechanisms compared to those at room temperature. Therefore, it is crucial to develop a novel, non-invasive, and contactless evaluation technique with high spatial resolution. In this study, we introduce an advanced photothermal coherence tomography method—frequency-multiplexed photothermal correlation tomography (FM-PCT)—to enable three-dimensional imaging of subsurface structures. A series of comprehensive experiments and analyses were conducted, including validation through comparison with X-ray micro-computed tomography (micro-CT) and multi-view tomographic evaluations of various (toughened) composite materials. The results demonstrate a significant reduction in the thermal diffusivity of CFRP laminates after LVI, with the reduction being more pronounced under low temperatures and higher impact energies. Moreover, among the studied composites, PA12/CF exhibits a distinct energy absorption mechanism, characterized by crack formation without delamination, in contrast to the damage modes observed in PEEK/CF, Neat/CF, and PA6.6/CF.

Limited by the Abbe diffraction limit and lateral thermal diffusion, thermograms in infrared thermography typically suffer from low spatial resolution. Although deep learning has demonstrated strong potential in enhancing resolution at minimal cost, the lack of large-scale, open-access training datasets specific to infrared thermography remains a significant challenge. To address this issue, we propose a novel transfer learning-based infrared super-resolution generative adversarial network (IR-SRGAN). This is, to the best of our knowledge, the first implementation of transfer learning in this context. An effective pre-trained model was first obtained using a large-scale dataset. Subsequently, the adversarial network and part of the generator were frozen, and a small-scale dataset of impact-damaged composite materials was fabricated to fine-tune the unfrozen components. Compared to directly training on either the source or target dataset, the transfer learning strategy prevents the model from being overly biased by the source dataset—which typically has higher resolution and scale—and simultaneously reduces computational cost in terms of both training time and GPU resources. We demonstrate the successful application of IR-SRGAN across multiple types of composite materials. Furthermore, the performance of IR-SRGAN was compared with conventional cubic interpolation and Enhanced Super-Resolution GAN (ESRGAN). The results show that IR-SRGAN effectively reconstructs natural textures in both damaged and undamaged regions, and significantly improves spatial resolution.

## CRediT authorship contribution statement

**Pengfei Zhu:** Data curation (equal), Investigation (equal), Writing – original draft (equal). **Hai Zhang:** Supervision (equal), Investigation (equal), Writing – revised draft (equal). **Stefano Sfarra:** Investigation (equal), Resources (equal), Writing – Reviewing & Editing (equal). **Fabrizio Sarasini:** Investigation (equal), Resources (equal), Writing – Reviewing & Editing (equal). **Clemente Ibarra-Castanedo:** Investigation (equal), Resources (equal), Writing – Reviewing & Editing (equal). **Xavier Maldague:** Supervision (equal).

## Declaration of competing interest

The authors declare that they have no known competing financial interests or personal relationships that could have appeared to influence the work reported in this paper.

## Data availability

Data will be made available on request.

## Acknowledgments

This work was supported by the Natural Sciences and Engineering Research Council (NSERC) Canada through the CREATE 'oN DuTy!' program (Grant no. 496439-2017) and the Canada Research Chair in Multipolar Infrared Vision (MiViM).